# Search for the Flavor-Changing Neutral-Current Decays $D^+ \to \pi^+\mu^+\mu^-$ and $D^+ \to \pi^+e^+e^-$

E. M. Aitala,[8] S. Amato,[1] J. C. Anjos,[1] J. A. Appel,[5] D. Ashery,[14] S. Banerjee,[5] I. Bediaga,[1] G. Blaylock,[2] S. B. Bracker,[15] P. R. Burchat,[13] R. A. Burnstein,[6] T. Carter,[5] H. S. Carvalho,[1] I. Costa,[1] L. M. Cremaldi,[8] C. Darling,[18] K. Denisenko,[5] A. Fernandez,[11] P. Gagnon,[2] S. Gerzon,[14] C. Gobel,[1] K. Gounder,[8] D. Granite,[7] A. M. Halling,[5] G. Herrera,[4] G. Hurvits,[14] C. James,[5] P. A. Kasper,[6] N. Kondakis,[10] S. Kwan,[5] D. C. Langs,[10] J. Leslie,[2] J. Lichtenstadt,[14] B. Lundberg,[5] A. Manacero,[5] S. MayTal-Beck,[14] B. Meadows,[3] J. R. T. de Mello Neto,[1] R. H. Milburn,[16] J. M. de Miranda,[1] A. Napier,[16] A. Nguyen,[7] A. B. d'Oliveira,[3,11] K. O'Shaughnessy,[2] K. C. Peng,[6] L. P. Perera,[3] M. V. Purohit,[12] B. Quinn,[8] S. Radeztsky,[17] A. Rafatian,[8] N. W. Reay,[7] J. J. Reidy,[8] A. C. dos Reis,[1] H. A. Rubin,[6] A. K. S. Santha,[3] A. F. S. Santoro,[1] A. J. Schwartz,[10] M. Sheaff,[17] R. A. Sidwell,[7] A. J. Slaughter,[18] J. G. Smith,[7] M. D. Sokoloff,[3] N. R. Stanton,[7] K. Sugano,[2] D. J. Summers,[8] S. Takach,[18] K. Thorne,[5] A. K. Tripathi,[9] S. Watanabe,[17] R. Weiss,[14] J. Wiener,[10] N. Witchey,[7] E. Wolin,[18] D. Yi,[8] R. Zaliznyak,[13] and C. Zhang[7]
(Fermilab E791 Collaboration)

[1] *Centro Brasileiro de Pesquisas Fisicas, Rio de Janeiro, Brazil*
[2] *University of California, Santa Cruz, California 95064*
[3] *University of Cincinnati, Cincinnati, Ohio 45221*
[4] *CINVESTAV, Mexico*
[5] *Fermilab, Batavia, Illinois 60510*
[6] *Illinois Institute of Technology, Chicago, Illinois 60616*
[7] *Kansas State University, Manhattan, Kansas 66506*
[8] *University of Mississippi, University, Mississippi 38677*
[9] *The Ohio State University, Columbus, Ohio 43210*
[10] *Princeton University, Princeton, New Jersey 08544*
[11] *Universidad Autonoma de Puebla, Mexico*
[12] *University of South Carolina, Columbia, South Carolina 29208*
[13] *Stanford University, Stanford, California 94305*
[14] *Tel Aviv University, Tel Aviv, Israel*
[15] *317 Belsize Drive, Toronto, Canada*
[16] *Tufts University, Medford, Massachusetts 02155*
[17] *University of Wisconsin, Madison, Wisconsin 53706*
[18] *Yale University, New Haven, Connecticut 06511*

ABSTRACT

We report the results of a search for the flavor-changing neutral-current decays $D^+ \to \pi^+\mu^+\mu^-$ and $D^+ \to \pi^+e^+e^-$ in data from Fermilab charm hadroproduction experiment E791. No signal above background is found, and we obtain upper limits on branching fractions, $B(D^+ \to \pi^+\mu^+\mu^-) < 1.8\times10^{-5}$ and $B(D^+ \to \pi^+e^+e^-) < 6.6 \times 10^{-5}$, at the 90% confidence level.

Flavor-changing neutral-current (FCNC) decays have played a major role in our understanding of weak decay processes. For example, the absence of such currents in neutral kaon decays in the strange quark sector helped motivate the Glashow-Iliopoulos-Maiani mechanism [1] for weak decay that predicted the existence of the charm quark before its experimental observation. In this Letter we present branching fraction limits from Fermilab experiment E791 for FCNC decays in the charm quark sector, $D^+ \to \pi^+\mu^+\mu^-$, $\pi^+e^+e^-$ [2]. These decays would proceed through an effective $c \to u$ quark transition. Such a transition is forbidden at the tree level within the Standard Model, but can occur through higher-order loop diagrams. The decays $D^+ \to \pi^+\ell^+\ell^-$ are thus sensitive to physics at high mass scales that cannot be accessed directly [3].

Within the Standard Model, the contribution of short-distance electroweak processes to $B(D^+ \to \pi^+\ell^+\ell^-)$ is expected to be less than $10^{-8}$ [4]; in contrast, the best published experimental limits are only 2 to $3 \times 10^{-3}$ ($\pi^+e^+e^-$) [5] and $2 \times 10^{-4}$ ($\pi^+\mu^+\mu^-$) [6]. Three-body FCNC decays such as $D^+ \to \pi^+\ell^+\ell^-$ are more sensitive to new physics involving vector or axial vector currents than are corresponding two-body decays such as $D^0 \to \ell^+\ell^-$ because three-body decays do not suffer from helicity suppression imposed by angular momentum conservation. In the search reported here, we find no evidence for FCNC $D^+ \to \pi^+\ell^+\ell^-$ decays. We set upper limits on the branching fractions that are an order of magnitude below those previously published.

The data were recorded from 500 GeV/$c$ $\pi^-$ interactions in five thin foils (one platinum, four diamond) separated by gaps of 1.34 to 1.39 cm. The E791 spectrometer is an upgraded version of the apparatus used in Fermilab charm experiments E516, E691, and E769 [7]. Momentum analysis is provided by two dipole magnets which bend particles in the horizontal plane. Twenty-three silicon microstrip detectors provide position information for track and vertex reconstruction.

Muon identification is obtained from a plane of 16 scintillation counters located behind shielding with a thickness equivalent to 2.5 meters (15 interaction lengths) of steel. These counters, 24 m from the target, are each 14 cm by 300 cm. They are mounted with their long sides horizontal and thus measure displacement in the vertical (nonbend) direction. The probability for misidentifying a hadron as a muon rises from $(4.5 \pm 0.5)\%$ for momenta greater than 20 GeV/$c$ to about 20% at 7 GeV/$c$. The efficiencies of the muon counters were measured to be $(99 \pm 1)\%$ in special runs using independent muon identification.

Electrons are identified by a calorimeter [8] of lead and liquid scintillator located 19 m from the target. The scintillator is divided into strips 3.17 cm in width [9], oriented transverse to the beam direction, with successive scintillator planes alternating among three stereo projections. Electron identification is based on energy deposition and transverse shower shape in the calorimeter. The probability that a pion is misidentified as an electron is 0.8%, nearly independent of momentum. Electron identification efficiency varies from 62% for momenta below 9 GeV/$c$ to 45% for momenta above 20 GeV/$c$.

The experiment recorded $2 \times 10^{10}$ events with a loose transverse energy trigger. After reconstruction, events with evidence of well-separated primary and secondary vertices were retained for further analysis. To search for $D^+$ FCNC decays in an unbiased way, the Cabibbo-suppressed decay $D^+ \to \pi^-\pi^+\pi^+$ was used to determine a single set of track and vertex selection criteria for all $D^+$ decays. The criteria are chosen to maximize $N_S/\sqrt{N_B}$, where $N_S$ and $N_B$ are the numbers of signal and background events, respectively, in the $\pi^-\pi^+\pi^+$ decay.

To separate charm candidates from background, we require the secondary vertex to be well-separated from the primary vertex and located well outside the target foils and other solid material, the momentum vector of the candidate $D^+$ to point back to the primary vertex, and the decay track candidates to pass closer to the secondary vertex than the primary vertex. Specifically, a three-prong secondary vertex must be separated by $> 20\,\sigma_L$ from the primary vertex and by $> 5.0\,\sigma_L$ from the closest material in the target foils, where $\sigma_L$ in each case is the calculated longitudinal resolution in the measured separation. The sum of the momentum vectors of the three tracks from this secondary vertex must miss the primary vertex by $< 40$ $\mu$m in the plane perpendicular to the beam. We form the ratio of each track's smallest distance from the secondary vertex to its smallest distance from the primary vertex, and require the product of these ratios to be less than 0.001. Finally, the net momentum of the $D^+$ candidate transverse to the line connecting the primary and secondary vertices must be less than 250 MeV/$c$.

The longitudinal and transverse position resolutions for the primary vertex are 350 $\mu$m and 6 $\mu$m, respectively. For three-prong secondary vertices from $D^+$ decays, the transverse resolution is about 9 $\mu$m, nearly independent of momentum; the longitudinal resolution is about 360 $\mu$m at 70 GeV/$c$ and increases roughly linearly with a slope of 30 $\mu$m per 10 GeV/$c$. Normalizing separation distances by their estimated resolutions reduces momentum-dependent effects.

The $K^-\pi^+\pi^+$ channel serves as the normalization mode for the FCNC branching fraction measurements; this mass spectrum, obtained with the selection criteria described above, is shown in Figure 1. A fit to a Gaussian signal and an exponential background gives 37,000 $D^+ \to K^-\pi^+\pi^+$ events.

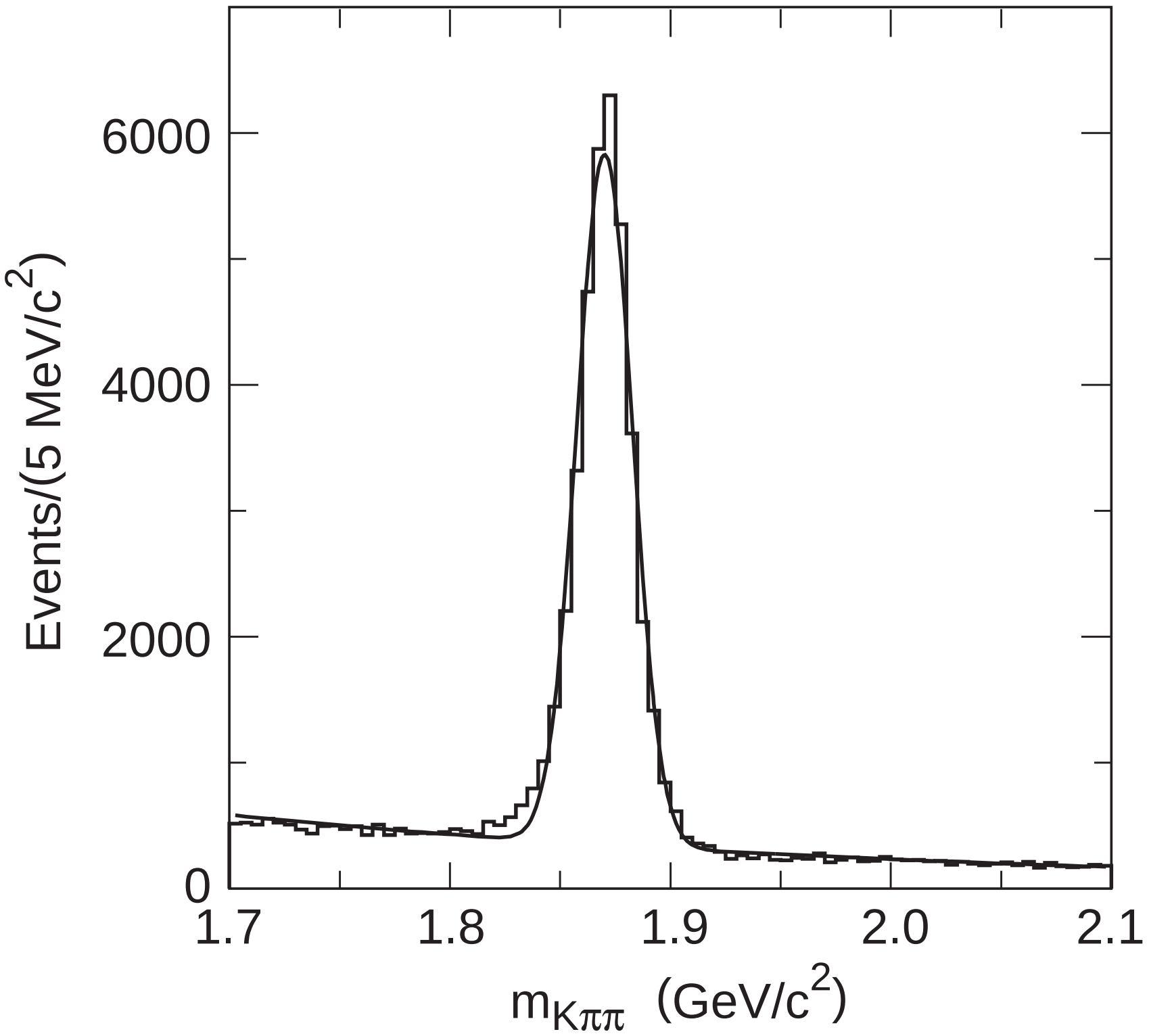

Figure 1: Mass spectrum of candidate $D^+ \to K^-\pi^+\pi^+$ events with final optimized selection criteria, used to provide the normalizing signal for FCNC analyses. The curve is a fit to a Gaussian signal and an exponential background.

---

The measured mass resolution of the $D^+$ peak is $(12 \pm 1)$ MeV/$c^2$.

In addition to meeting the selection criteria described above, candidates for $D^+ \to \pi^+\mu^+\mu^-$ are required to have two oppositely-charged tracks from the candidate decay vertex identified as muons. Since the momenta and directions of charged tracks are measured in tracking devices upstream of absorbing material in the muon system, multiple scattering in this shielding introduces a significant uncertainty at the scintillation counters in predicted positions of low-momentum tracks. To identify muons efficiently, we require a signal from a scintillation counter whose closest edge is within 1.0 $\sigma_{MS}$ from the track's projected vertical position, where $\sigma_{MS}$ is the predicted root-mean-square position error due to multiple scattering for a muon of the measured momentum. Muon candidates must have momenta greater than 7 GeV/$c$, and the two muon candidates must be identified by separate scintillation counters. The latter requirement reduces background by more than 50% and, for a uniformly populated Dalitz plot, reduces the dimuon efficiency by only 14%. A major source of candidates with $\pi^+\mu^+\mu^-$ invariant mass below

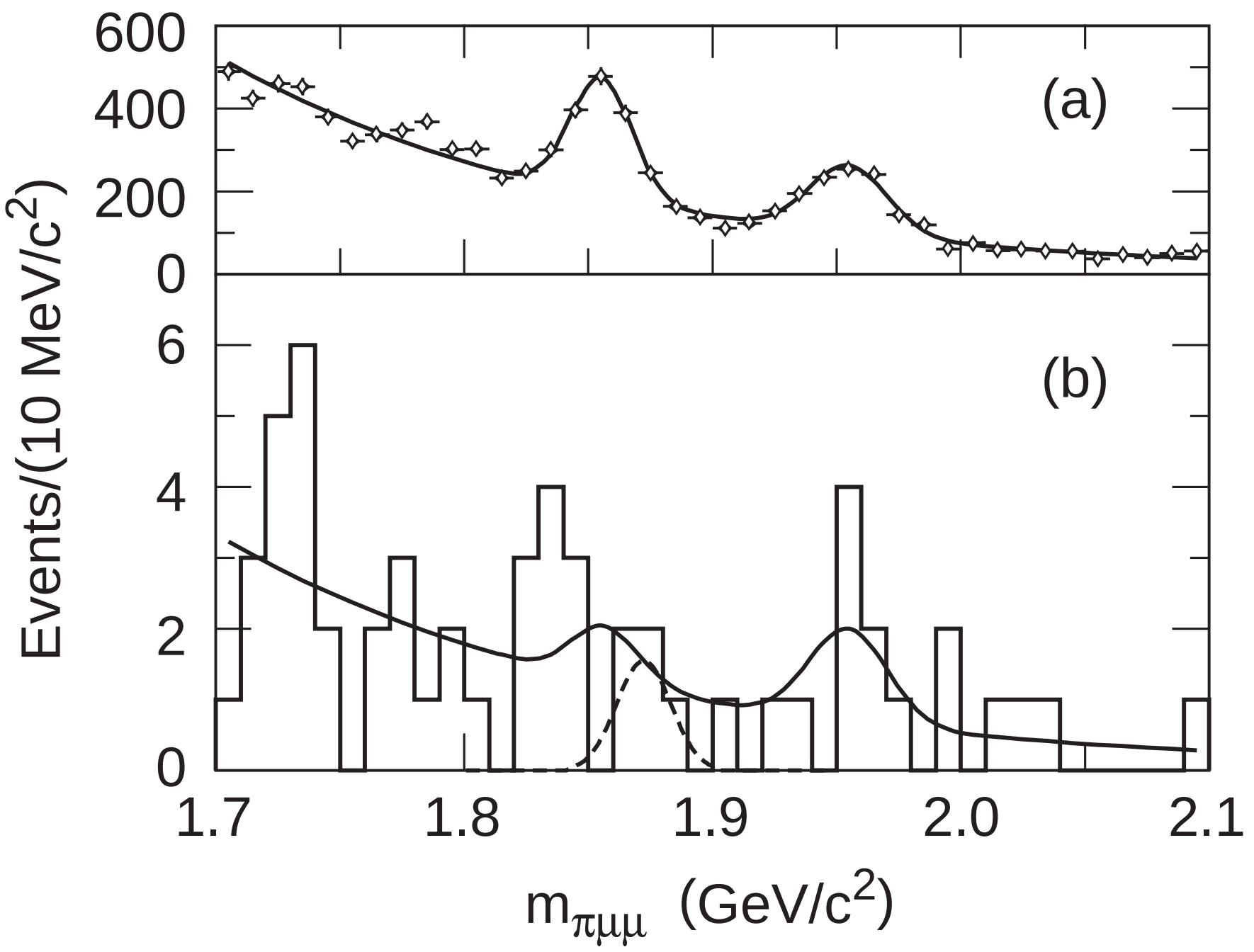

Figure 2: Search for a $D^+ \to \pi^+\mu^+\mu^-$ signal. (a) Invariant mass spectrum under a $\pi^+\mu^+\mu^-$ hypothesis but with no muon identification requirement (diamonds). The curve, which is a fit to the sum of Gaussian peaks from misidentified $D^+ \to \pi^-\pi^+\pi^+$ and $D_s^+ \to \pi^-\pi^+\pi^+$ and an exponential background, determines the central values and widths of the peaks. (b) Invariant $\pi^+\mu^+\mu^-$ mass spectrum for events with muon identification (histogram). The solid curve is the best fit to a sum of contributions from $D^+ \to \pi^+\mu^+\mu^-$, feedthrough from $D^+ \to \pi^-\pi^+\pi^+$ and $D_s^+ \to \pi^-\pi^+\pi^+$, and an exponential background. The dashed curve shows the size and shape of the $D^+ \to \pi^+\mu^+\mu^-$ contribution ruled out at 90% CL.

---

that of the $D^+$ is misidentified $K^-\pi^+\pi^+$ decays. We exclude candidates with $K^-\pi^+\pi^+$ invariant mass inside the interval 1.850 to 1.890 GeV/$c^2$. This requirement has negligible effect on the $D^+ \to \pi^+\mu^+\mu^-$ efficiency but makes fitting the background more straightforward.

In the search for a $D^+ \to \pi^+\mu^+\mu^-$ signal it is important to account correctly for feedthrough from misidentified $D^+ \to \pi^-\pi^+\pi^+$ and $D_s^+ \to \pi^-\pi^+\pi^+$ decays. The procedure is illustrated in Figure 2. First, the invariant mass spectrum, under a $\pi^+\mu^+\mu^-$ hypothesis but with no muon identification requirement (Figure 2a), is fit to the sum of an exponentially-falling distribution and two Gaussians describing the $D^+ \to \pi^-\pi^+\pi^+$ and $D_s^+ \to \pi^-\pi^+\pi^+$ peaks (solid curve). Because of the incorrectly-assigned daughter masses, these $\pi^-\pi^+\pi^+$ signals are broadened and shifted downward by about 15 MeV/$c^2$ from the true parent masses. The central values and widths of these peaks are retained for the fit described below. Next, the muon identification requirement is imposed, giving the spectrum shown by the histogram in Figure 2b. We then search for an FCNC signal with a binned maximum likelihood fit with four components: a Gaussian centered at the $D^+$ mass for $D^+ \to \pi^+\mu^+\mu^-$, two Gaussians describing feedthrough from $D^+ \to \pi^-\pi^+\pi^+$ and $D_s^+ \to \pi^-\pi^+\pi^+$, and an exponential distribution. The central values and widths of the feedthrough peaks are constrained to the values from the fit in Figure 2a, and the width of the signal Gaussian is fixed at its observed value in $D^+ \to \pi^-\pi^+\pi^+$ decays, 11 MeV/$c^2$. The sizes of the four contributions are allowed to vary independently. The result of this fit is shown by the solid curve in Figure 2b. We find $0.35^{+3.04}_{-2.47}$ events from $D^+ \to \pi^+\mu^+\mu^-$. The 90% confidence level (CL) upper limit is 4.36 events, determined by the point at which the log likelihood function falls below its maximum value by $(1.28)^2/2$. The size and shape of the signal excluded at 90% CL is shown by the dashed curve in Figure 2b.

Variations on this technique give consistent results. Constraining the relative amounts of $D^+ \to \pi^-\pi^+\pi^+$ and $D_s^+ \to \pi^-\pi^+\pi^+$ feedthrough to be the same as in Figure 2a leads to an upper limit of 3.2 events at 90% CL, while use of a simple mass window instead of a maximum likelihood fit gives 4.5 events. We have also tested the procedure with ensembles of simulated experiments in which fixed numbers (2 - 10) of simulated FCNC signal events, drawn randomly from a Gaussian mass distribution, are added to the observed spectrum and successfully found by the fit.

The 90% CL upper limit on the branching fraction $B$ for $D^+ \to \pi^+\mu^+\mu^-$ is given by

$$B(D^+ \to \pi^+\mu^+\mu^-) < \frac{U(\pi^+\mu^+\mu^-)}{N(K^-\pi^+\pi^+)} \frac{1}{\eta_\mu} f_{SYS}\, B(D^+ \to K^-\pi^+\pi^+)$$

where $U(\pi^+\mu^+\mu^-) = 4.36$ is the 90% CL upper limit on the number of signal events, $N(K^-\pi^+\pi^+) = 37,006 \pm 204$ is the number of events in the normalizing channel, $\eta_\mu = 0.625 \pm 0.075$ is the relative efficiency of the $\pi^+\mu^+\mu^-$ and $K^-\pi^+\pi^+$ channels, $f_{SYS} = 1.03$ is the factor [10] by which the limit is degraded by systematic uncertainties, and $B(D^+ \to K^-\pi^+\pi^+) = (9.1 \pm 0.6)\%$ [11].

The relative efficiency $\eta_\mu$ is determined primarily from a Monte Carlo calculation. It includes factors arising from different decay kinematics (1.12), the 7 GeV/$c$ muon track momentum requirement (0.96), the muon identification (0.67) and separate-counter (0.89) requirements, and the relative trigger efficiency (0.98). The systematic uncertainties which give rise to $f_{SYS}$, totalling 15% when added in quadrature, include those in $N(K^-\pi^+\pi^+)$, $\eta_\mu$ and $B(D^+ \to K^-\pi^+\pi^+)$. The factor $f_{SYS}$ is then calculated by the prescription of Ref. [10].

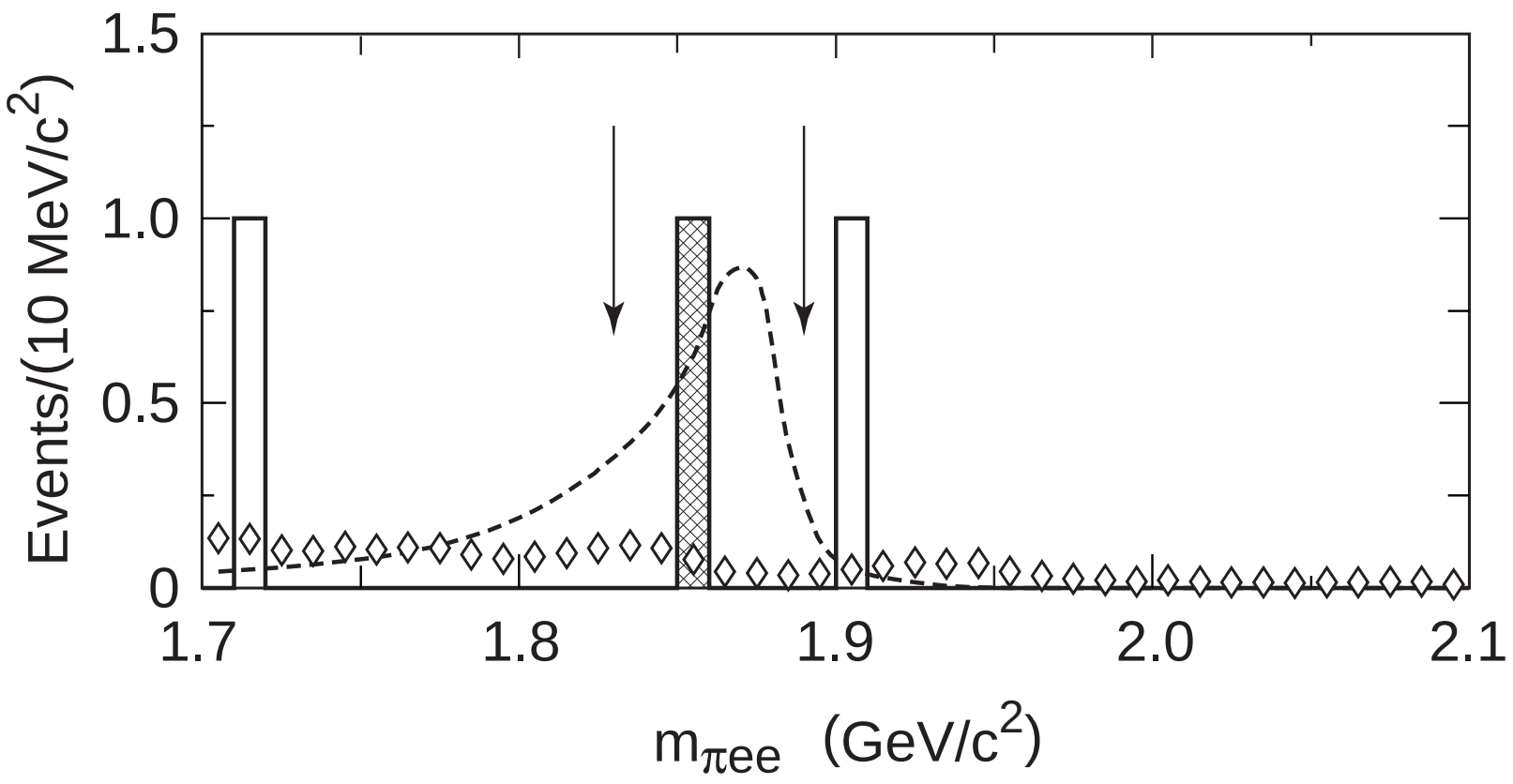

Figure 3: Search for a $D^+ \to \pi^+ e^+ e^-$ signal: invariant mass spectrum with $\pi^+ e^+ e^-$ hypothesis. Three events pass the electron identification requirement (histogram). One of them lies in the signal region between the arrows. Background is estimated from the $\pi^+ e^+ e^-$ invariant mass spectrum without electron identification requirement, normalized to two events outside the signal region, as shown by the diamond points. The dashed curve shows the size and shape of the bremstrahllung-broadened $D^+ \to \pi^+ e^+ e^-$ signal ruled out at 90% CL.

---

The resulting upper limit on the branching fraction is $B(D^+ \to \pi^+ \mu^+ \mu^-) < 1.8 \times 10^{-5}$ at 90% CL. The limit is quite stable under variation of the vertex selection criteria and of the muon momentum requirements. To facilitate combining our result with those of other experiments, it can be expressed as a branching fraction measurement: $B(D^+ \to \pi^+ \mu^+ \mu^-) = (0.14^{+1.2}_{-1.0}) \times 10^{-5}$, where the quoted errors are statistical only; the systematic errors, approximately 15% of $B$, are negligible by comparison.

Monte Carlo calculations show that the experimental acceptance is nearly uniform across the $\pi^+ \mu^+ \mu^-$ Dalitz plot for squared dimuon mass $m^2_{\mu\mu}$ greater than 0.8 $(\mathrm{GeV}/c^2)^2$. For smaller $m^2_{\mu\mu}$ the acceptance decreases roughly linearly to near zero at minimum $m^2_{\mu\mu}$ [12].

The search for $D^+ \to \pi^+ e^+ e^-$ employs vertex selection and $D^+ \to K^- \pi^+ \pi^+$ exclusion criteria identical to those for $\pi^+ \mu^+ \mu^-$. Candidate $D^+ \to \pi^+ e^+ e^-$ decays contain two oppositely-charged tracks identified as electrons and have $\pi^+ e^+ e^-$ invariant mass consistent with $D^+$. Because there are only three identified $\pi^+ e^+ e^-$ candidates, we use the simple search-window procedure illustrated in Figure 3. The $\pi^+ e^+ e^-$ events are shown as a solid histogram. One of them (shaded) is inside the search window (indicated by arrows) of 1.830 to 1.890 $\mathrm{GeV}/c^2$. The $\pi^+ e^+ e^-$ spectrum without electron identification is normalized to two events outside the search window, and then used to predict the expected number inside the window: $0.42 \pm 0.29$

events. The signal is thus $0.58 \pm 1.04$ events. Using the method for Poisson processes with background described in Reference [11], we find the 90% CL upper limit for a signal of one event with 0.42 expected background events to be 3.56. The size and shape of the signal excluded at 90% CL is shown by the dashed curve.

The upper limit on $B(D^+ \to \pi^+e^+e^-)$ is calculated from

$$B(D^+ \to \pi^+e^+e^-) < \frac{U(\pi^+e^+e^-)}{N(K^-\pi^+\pi^+)}\frac{1}{\eta_e} f_{SYS} B(D^+ \to K^-\pi^+\pi^+)$$

where $U(\pi^+e^+e^-) = 3.56$ events at 90% CL, $N(K^-\pi^+\pi^+) = 37,006 \pm 204$ events in the normalizing channel, $\eta_e = 0.138 \pm 0.021$, and $f_{SYS} = 1.04$. The relative efficiency $\eta_e$ of $\pi^+e^+e^-$ and $K^-\pi^+\pi^+$, determined from a Monte Carlo calculation, is dominated by the dielectron detection efficiency (0.32) and by the exclusion of much of the expected long bremsstrahlung tail in a $D^+ \to \pi^+e^+e^-$ mass peak from the mass window (0.66). The systematic uncertainties which give rise to $f_{SYS}$ [10], totalling 16% when added in quadrature, include those in $N(K^-\pi^+\pi^+)$, $\eta_e$ and $B(D^+ \to K^-\pi^+\pi^+)$.

The resulting upper limit on the branching fraction is $B(D^+ \to \pi^+e^+e^-) < 6.6\times 10^{-5}$ at 90% CL. Alternatively, the branching fraction with one standard deviation errors is $B(D^+ \to \pi^+e^+e^-) = (1.0 \pm 1.9) \times 10^{-5}$. The efficiency across the Dalitz plot is approximately uniform for squared dielectron masses $m_{ee}^2$ above about 0.4 $(\mathrm{GeV}/c^2)^2$, but decreases roughly linearly to near zero at minimum $m_{ee}$ [12].

In summary, Fermilab experiment E791 has obtained new upper limits on the branching fractions $B$ for the three-body flavor-changing neutral-current decays $D^+ \to \pi^+\mu^+\mu^-$ and $D^+ \to \pi^+e^+e^-$ that are an order of magnitude smaller than those previously published. At the 90% CL $B(D^+ \to \pi^+\mu^+\mu^-) < 1.8 \times 10^{-5}$, and $B(D^+ \to \pi^+e^+e^-) < 6.6 \times 10^{-5}$.

We gratefully acknowledge the assistance of the staffs of Fermilab and of all the participating institutions. This research was supported by the Brazilian Conselho Nacional de Desenvolvimento Científico e Technológico, CONACyT (Mexico), the Israeli Academy of Sciences and Humanities, the Fulbright Foundation, the U.S. Department of Energy, the U.S.-Israel Binational Science Foundation, and the U.S. National Science Foundation. Fermilab is operated by the Universities Research Association, Inc., under contract with the United States Department of Energy.